\documentclass[twocolumn]{aastex631}

\usepackage{CJK}
\usepackage{placeins}
\usepackage{amsmath}
\usepackage{cleveref}
\usepackage{soul}

\newcommand{\m}{{\rm ~m}}
\newcommand{\km}{{\rm ~km}}
\newcommand{\au}{{\rm ~au}}

\newcommand{\s}{{\rm ~s}}
\newcommand{\h}{{\rm ~h}}

\newcommand{\J}{{\rm ~J}}
\newcommand{\K}{{\rm ~K}}

\newcommand{\YORP}{{\rm YORP}}
\newcommand{\B}{{\rm B}}

\newcommand{\BYORP}{{\rm B}}

\newcommand{\tide}{{\rm t}}
\newcommand{\Y}{{\rm Y}}
\newcommand{\YS}{{\rm YS}}

\newcommand{\rms}{{\rm s}}
\newcommand{\p}{{\rm p}}
\newcommand{\Roche}{{\rm Roche}}

\newcommand{\rmh}{{\rm h}}
\newcommand{\syn}{{\rm syn}}

\shorttitle{The binary Yarkovsky effect}
\shortauthors{Wen-Han Zhou et al}

\begin{document}
\begin{CJK*}{UTF8}{bsmi}

\title{The Yarkovsky effect on the long-term evolution of binary asteroids}

\author[0000-0003-4229-8936]{Wen-Han Zhou (周文翰)}
\affiliation{Universit\'e C\^ote d'Azur, Observatoire de la C\^ote d'Azur, CNRS, Laboratoire Lagrange, Nice, France}

\author[0000-0002-6034-5452]{David Vokrouhlick\'y}
\affiliation{Astronomical Institute, Charles University, V Hole\v{s}ovi\v{c}k\'ach 2,
             CZ 18000, Prague 8, Czech Republic}

\author[0000-0002-2533-3077]{Masanori Kanamaru}
\affiliation{Department of Earth and Planetary Science, School of Science, the University of Tokyo, Japan}

\author[0000-0002-3544-298X]{Harrison Agrusa}
\affiliation{Universit\'e C\^ote d'Azur, Observatoire de la C\^ote d'Azur, CNRS, Laboratoire Lagrange, Nice, France}

\author[0000-0001-8434-9776]{Petr Pravec}
\affiliation{Astronomical Institute, Astronomical Institute of Czech Academy, Ond\v{r}ejov, CZ-25165, Czech Republic}

\author[0000-0002-8963-2404]{Marco Delbo}
\affiliation{Universit\'e C\^ote d'Azur, Observatoire de la C\^ote d'Azur, CNRS, Laboratoire Lagrange, Nice, France}

\author[0000-0002-0884-1993]{Patrick Michel}
\affiliation{Universit\'e C\^ote d'Azur, Observatoire de la C\^ote d'Azur, CNRS, Laboratoire Lagrange, Nice, France}
\affiliation{The University of Tokyo, Department of Systems Innovation, School of Engineering, Tokyo, Japan}

\begin{abstract}

We explore the Yarkovsky effect on small binary asteroids. While significant attention has been given to the binary YORP effect, the Yarkovsky effect is often overlooked. We develop an analytical model for the binary Yarkovsky effect, considering both the Yarkovsky-Schach and planetary Yarkovsky components, and verify it against thermophysical numerical simulations. {We find that the Yarkovsky force could change the mutual orbit when the asteroid's spin period is unequal to the orbital period.} Our analysis predicts new evolutionary paths for binaries. For a prograde asynchronous secondary, the Yarkovsky force will migrate the satellite towards the location of the synchronous orbit on ${\sim} 100$ kyr timescales, which {could be faster than other synchronization processes such as YORP and tides.} For retrograde secondaries, the Yarkovsky force always migrates the secondary outwards, which could produce asteroid pairs with opposite spin poles. Satellites spinning faster than the Roche limit orbit period {(e.g. from $\sim$4~h to $\sim$10~h)} will migrate inwards until they disrupt, reshape, or form a contact binary. We also predict a short-lived equilibrium state for asynchronous secondaries where the Yarkovsky force is balanced by tides. We provide calculations of the Yarkovsky-induced drift rate for known asynchronous binaries. If the NASA DART impact broke Dimorphos from synchronous rotation, we predict that Dimorphos's orbit will shrink by $\dot a\, {\sim}\, 7$ cm~yr$^{-1}$, which can be measured by the Hera mission. We also speculate that the Yarkovsky force may have synchronized the Dinkinesh-Selam system after a possible merger of Selam's two lobes.

\end{abstract}

\section{Introduction} \label{sec:intro}

Binary asteroids are found throughout the Solar System at a wide range of size scales. Their formation mechanisms are also diverse. Km-sized systems are generally thought to form by rotational disruption of the primary resulting from radiative torques \citep[e.g.,][]{Walsh2008}, large main-belt systems are thought to form by collisions \citep[e.g.,][]{Michel2001,Durda2004}, while binaries in the Kuiper Belt are thought to be primordial, forming directly from the streaming instability \citep[e.g.,][]{Nesvorny2010}. This study primarily focuses on ${\sim}$km-sized binaries found among both the near-Earth asteroids (NEAs) and main-belt asteroids (MBAs). These systems are small and close enough to the Sun that radiation forces play an important role in their long-term evolution. Understanding their long-term dynamics is crucial to trace back their evolution and estimate their lifetime, which also provides information on physical properties and geologic structures of asteroids.

It is widely accepted that the long-term dynamics of binaries are dominated by tides and the Binary YORP (BYORP) effect, which is a radiative torque that modifies the orbit of the secondary asteroid \citep{Cuk2005,vetal15}. Tidal dissipation can either drive the secondary outward or inward, depending on whether the secondary's mean motion is slower or faster than the primary's spin \citep{Murray1999}. The primaries of binary NEAs typically have short rotation periods, in the range 2.2-4.5 h \citep{WalshJacobson2015}, which is likely due to the formation of the system by rotational failure \citep{Pravec2007}. For simplicity, we assume the primary's spin rate always exceeds the secondary's mean motion and that tides will consequently drive the secondary outward. For small eccentricities, the time evolution of the binary semimajor axis can be written as \citep{Murray1999}
\begin{equation}
\label{eq:a_tide}
    \dot a_\tide = 3 \frac{k_\p}{Q_\p}\frac{m_\rms}{m_\p} \left( \frac{{r_\p}}{{a}} \right)^5 n a.
\end{equation}
Here $k_p$, $Q_\p$ and $m_\p$ are the tidal Love number, quality factor and mass of the primary, while $n$, $a$ and $m_\rms$ are the mean motion, semimajor axis and the mass of the secondary, respectively. Throughout this manuscript, the subscript ''p`` denotes the primary while the subscript ''s`` denotes the secondary. The nomenclature and symbols are given in Table~\ref{tab:Nomenclature}.

While $\dot a_\tide$ decreases dramatically with the semimajor axis ($\dot a_\tide \propto a^{-11/2}$), the drift rate caused by the BYORP effect becomes greater with the increasing semimajor axis \citep{Cuk2005, Jacobson2011b, vetal15}. The averaged semimajor axis drift rate under BYORP is 
\begin{equation}
    \dot a_\BYORP = \frac{{2 f_\BYORP \mathcal{F}}}{{n}}.
\end{equation}
Here, $f_\BYORP$ is the dimensionless BYORP coefficient that can be positive or negative, depending on the shape and surface morphology of the secondary\footnote{In the literature, the BYORP coefficient is often referred to as B but we denote it as $f_\BYORP$ to maintain consistency with other coefficients in this paper.}. The calculated absolute value of $f_\BYORP$ for polyhedron asteroid models shows a large range from $10^{-4}$ to $10^{-1}$ \citep{Steinberg2011} with a typical value of $10^{-3}$ \citep{Jacobson2011b}.
The nominal radiation pressure per unit mass $\mathcal{F}$ is defined as
\begin{equation}
    \mathcal{F} = \frac{\Phi (1 - A)\,\pi r_\rms^2}{m_\rms c}.
\end{equation}
where $\Phi = 1364\, (a_{\rm h} / \rm au)^{-2}$~W~m$^{-2}$ is the solar flux, $A$ is the Bond albedo and $c$ is the speed of light. Here $a_{\rm h}$ is the heliocentric orbital semimajor axis of the binary system. The BYORP effect could drive the secondary either outward to an unstable orbit, where external gravitational perturbations would finally destroy the binary system \citep{Cuk2007}, or inward until the secondary gets tidally disrupted or the BYORP effect is balanced by the tidal effect \citep{Jacobson2011b}. The theoretical timescale of the BYORP effect for NEAs is short \citep[e.g., $\leq 10^5 $~yr,][]{Cuk2005, Cuk2007} compared to their dynamical lifetime \citep[e.g., $\sim 10^7 $~yr,][]{Gladman2000}, indicating that the observed binary asteroids are either very young or old enough, if reached a BYORP-tide equilibrium. However, available measurements suggest that binary systems are evolving at much lower rates than predicted by BYORP. The binary asteroid system 1996~FG$_3$ is observed to have a semimajor axis drift of $-0.07 \pm 0.34~\rm cm$~yr$^{-1}$ \citep{Scheirich2015}, which is much lower than the predicted values of $2.3~$cm~yr$^{-1}$ \citep{Scheirich2015} or 7~cm~yr$^{-1}$ \citep{Mcmahon2010} based on the secondary shape model. Similarly, based on an available shape model of the secondary, the binary system 1999~KW$_4$ has been estimated to have a BYORP drift rate of 6.98~cm~yr$^{-1}$ \citep{Mcmahon2010} or $8.53$~cm~yr$^{-1}$ \citep{Scheirich2021}, while observations on the mutual orbit reports an outward drift rate of 1.2~cm~yr$^{-1}$ corresponding to $f_\BYORP \sim 0.00157$ if tides are neglected \citep{Scheirich2021}. The two orbital solutions for the binary system 2001~SL$_9$ have drift rates of $\dot a$ of $-2.8\pm 0.2$~cm~yr$^{-1}$ or $-5.1\pm 0.2$~cm~yr$^{-1}$ corresponding to $f_\BYORP = 0.006$ or $0.01$, respectively \citep{Scheirich2021}. Since there is no available shape model for the secondary in 2001~SL$_9$, a theoretical value of $f_\BYORP$ cannot be derived. {The observation data of Didymos-Dimorphos system before the DART impact show a small drifting rate of $-0.08 \pm 0.02~$cm~yr$^{-1}$\citep{Scheirich2022, Scheirich2024, Naidu2024}.}

Several mechanisms have been proposed for weakening the BYORP effect. The BYORP torque can be weakened, or even removed, by either non-synchronous state of the satellite \citep{Cuk2005} or its non-principal axis (NPA) rotation \citep{Quillen2022}. Another possibility is a rotational state referred to as the ``barrel instability'' \citep{Cuk2021}, in which the satellite rolls about its longest axis during its orbital motion and its longest axis remains approximately aligned towards the primary \citep{Agrusa2021}. A recent study also suggests that the BYORP coefficient can be reduced by an order of magnitude for satellites like Dimorphos, the secondary of the binary asteroid Didymos, which has an overall ``smooth'' shape, made up of boulders that are all significantly smaller than the size of the body \citep{Cuk2023}.

In this work, we investigate the Yarkovsky effect that has been largely overlooked in the context of the long-term evolution of binary asteroids. The Yarkovsky effect, which is the radiation force raised on the afternoon side of a rotating object, has been well studied for single asteroids \citep{Vokrouhlicky1998, Vokrouhlicky1999, Bottke06, vetal15}. However, its impact on binary asteroids remains less explored. The Yarkovsky effect on a binary consists of two components: the Yarkovsky-Schach (YS) effect and the planetary Yarkovsky effect. The YS effect is caused by: (1) elimination of the satellite irradiation by sunlight when it is located in the primary shadow; and (2) the related asymmetric thermal cooling and heating of the secondary after it enters and exits the shadow  (in fact, there is also a similar effect on the primary related by the shadow of the secondary, but this produces smaller dynamical perturbation). The YS effect has been studied for Earth satellites \citep{Rubincam1982, Rubincam1987, Milani1987,Farinella1996}, space debris \citep{Murawiecka2018}, and Saturn's rings \citep{rub2006, Vokrouhlicky2007}. This effect was noticed for binary asteroids too \citep{Vokrouhlicky2005}, but not studied in detail yet. The planetary Yarkovsky effect is simply the Yarkovsky effect caused by the primary's radiation instead of the Sun \citep{rub2006,Vokrouhlicky2007}.

In this paper, we describe the binary Yarkovsky effect in Sec.~\ref{sec2} and discuss its implications on the long-term evolution of binary systems in Sec.~\ref{sec3}. The main results are summarized in Sec.~\ref{sec4}.

\begin{figure}
    \centering
     \includegraphics[width=0.5\textwidth]{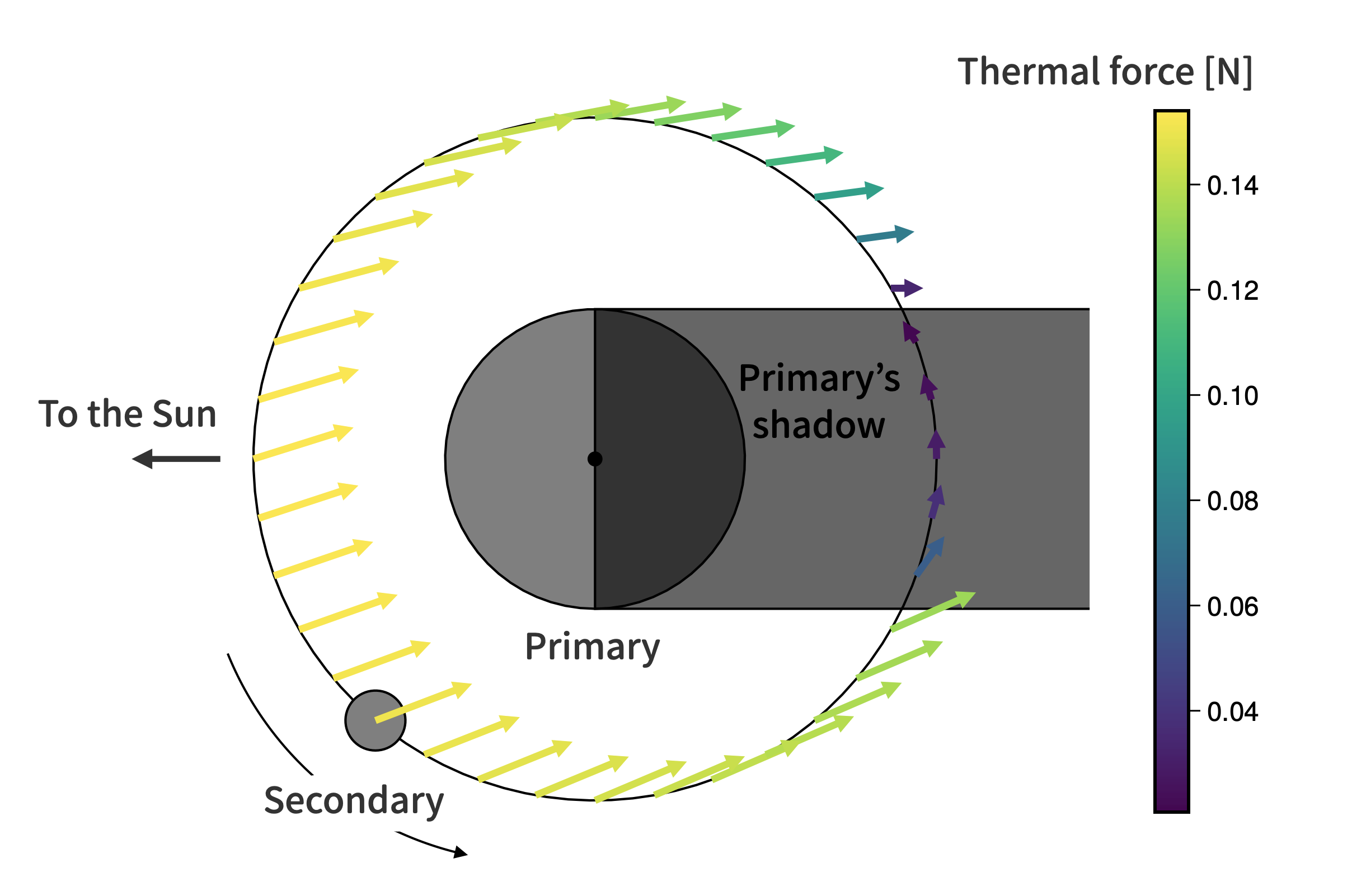}
     \caption{The principle of the Yarkovsky-Schach (YS) effect. A binary system consists of a larger primary and a smaller secondary (satellite). The relative orbit is assumed circular, the satellite has a zero obliquity and rotation synchronous with the motion about the primary; both periods are much smaller than the period of the binary heliocentric motion, such that during one satellite orbit about the primary the Sun is assumed fixed and in the orbital plane of the satellite. The color-coded arrows attached to the satellite represent its thermal acceleration due to solar irradiation (see also the side bar); the tilt away from the opposite direction to the Sun is due to the satellite thermal inertia. The specific values were computed using the numerical model and binary parameters from Sec.~\ref{nums}. In absence of the satellite passage through the primary's shadow, the thermal acceleration would be constant. The orbit-averaged effect on
   the satellite distance from the primary would be zero. The  essence of the YS effect is due to satellite crossing the primary's shadow. The interrupted solar irradiation results in the satellite cooling such that the thermal acceleration drops and tilts. Upon leaving the shadow the satellite heats, slowly regaining the thermal state at the subsolar configuration. The net
   budget of the transverse component of the thermal acceleration may be non-zero, depending on the satellite rotation rate and obliquity, resulting in a secular change of its distance from the primary.}
       \label{fig:thermal_force}
   \end{figure}

\section{Theory} \label{sec2}

\subsection{Analytical model}

When the secondary enters the shadow of the primary, its surface temperature drops, leading to a reduced Yarkovsky force. After the secondary exits the shadow of the primary, its temperature increases, restoring the Yarkovsky force level before entering the shadow. However, these two processes are not exactly balanced, resulting in a net perturbation over the orbit which leads to a secular change of $a$. This is the basis of the YS effect, whose concept is displayed in Fig.~\ref{fig:thermal_force}. The necessary condition for the YS effect to operate is therefore that the secondary enters the shadow of the primary. This constrains the inclination $i$ between the orbital plane defined by the secondary motion about the primary and the orbital plane of the binary barycenter about the Sun: $i < r_\p / a$ 
implying the satellite crosses the shadow in every orbit about the primary. A non-zero inclination could {weaken} the YS effect as the time fraction in the shadow {decreases} with the inclination \citep{Murawiecka2018}. However, a larger inclination $i > r_\p /a $ will result in only a fraction of the heliocentric orbit where the secondary can undergo an eclipse and therefore lead to a weakened YS effect. {In fact, binary systems that have been discovered tend to exhibit a preferred inclination of approximately $0\circ$ or $180\circ$ \citep{Pravec2012}.} Since the YORP effect drives the primary's obliquity to $0^\circ$ or $180^\circ$ \citep{Rubincam00}, the optimum condition $i<r_\p/a$ is usually easily satisfied for small binaries if they form via YORP-driven spin up and mass shedding followed by reaccumulation in the equatorial plane of the primary \citep{Walsh2008, Pravec2012, Agrusa2024} For simplicity, in the following, we take $i=0^\circ$. We first develop a simple analytical YS effect model, and later in this section, we justify it by comparison with the results of a numerical simulation.

There is also a ``mirror'' YS effect related to the satellite shadow which perturbs the thermal state of the primary. In principle, the corresponding drift rate of the satellite semimajor axis $\dot a$ may be described by a similar approach used for the core YS effect on the satellite. While algebraic complications would arise due to primary's larger size than the satellite's cross-section, it is
conceivable that the primary-driven YS component would be by a factor $ \sim (r_{\rm s}/r_{\rm p})^2$ smaller than the secondary-driven YS effect. As our ambition is to provide a simple and introductory analytical estimate of the YS effect, we neglect the thermal acceleration of the primary at this moment. 

Returning to the analytical formulation of the YS effect for the satellite, we assume that both the primary and the secondary have a spherical shape with radii  $r_\p $ and $r_\rms$, respectively (non-sphericity of both components may result in corrections, which are typically lower than the aimed accuracy of our simple analytical model). The heliocentric orbit of the barycenter, and the relative orbit of the two components in the binary, are both assumed to be circular. We denote the semimajor axis of the secondary orbit about the primary by $a$, the corresponding mean motion $n$, and the spin rate of secondary $\omega$. We introduce the frequency ratio $m = |\omega / (n - n_{\rm h})| \simeq  |\omega / n|$, where $n_{\rm h}$ is the heliocentric mean motion of the binary system. In this work
we assume $n_{\rm h}\ll n$, thus $n - n_{\rm h}\simeq n$ in the denominator of $m$.
We assume the secondary is in principal axis rotation. Complex rotational states such as a tumbling state or the so-called barrel instability are left for future investigation. 

The complete mathematical solution of the YS-effect for a small satellite orbiting a large primary is given in \citet{Vokrouhlicky2007} (a ring particle about Saturn, in their context). The semimajor axis drift rate of the secondary due to the YS-effect has a generic form
\begin{equation}
\label{eq:a_YS}
    \dot a_{\YS} =  \frac{2 f_{\rm YS} \mathcal{F}}{n}.
\end{equation}
The dimensionless coefficient $f_{\rm YS}$ is called the YS coefficient in this work, and depends on the physical properties of the binary system, such as the mutual orbital period, sizes of the two bodies composing the binary, and thermal properties. In fact, $f_\YS$ is the sum of the diurnal component and the seasonal component:
\begin{equation}
    f_\YS = f_{\YS,{\rm d}} + f_{\YS,{\rm s}} ,
\end{equation}
where
\begin{align}
    f_{\YS,{\rm d}} &= \phantom{-}{4 c_1 \over 9} \left[V(z_{m-1}) \cos^4{\varepsilon \over 2} - V(z_{m + 1})\sin^4{\varepsilon \over 2}\right], \label{eq:f_YS_d}\\
    f_{\YS,{\rm s}}&= - {2 c_1 \over 9} V(z_1)\,\sin^2{\varepsilon} \label{eq:f_YS_s}.
\end{align}
Here $z_{ m \pm 1 } = \sqrt{-{\imath }(m\pm1)}\,r_\rms / l_n$ ($\imath=\sqrt{-1}$), $c_1 \simeq r_\p / \pi a$ expresses orbital fraction spent by the satellite in the primary's shadow, and $V(z)$ a real-value function defined by 
\begin{equation}
    V(z) =  \operatorname{Im}{\left( 1 + \chi\, {z \over j_1(z)} {{\rm d}j_1(z) \over {\rm d}z}\right)^{-1}}
\end{equation}
with $j_1(z)$ denoting the spherical Bessel function of the first kind and order $1$
\begin{equation}
    j_1(z) = {\sin z \over z^2} - {\cos z \over z}.
\end{equation}
The thermal penetration depth $l_n$ at the satellite mean motion frequency $n$ is given by $l_n = \sqrt{K_\rms/(\rho_\rms C_\rms n) }$, where $K_\rms$ is the thermal conductivity, $C_\rms$ is the heat capacity and $\rho_\rms$ is the surface density of the satellite. The variable $\chi$ is defined as
\begin{equation}
    \chi = {K_\rms  \over \sqrt{2} r_\rms \epsilon \sigma T_{\rm sub}^3 c_0^{3/4}},
\end{equation}
with $c_0 = 1 - c_1$ and the subsolar temperature $T_{\rm sub}$ defined by
$\epsilon \sigma T_{\rm sub}^4 = (1-A)\,\Phi$, $\epsilon$ the thermal emissivity and $\sigma$ the Stefan-Boltzmann constant. Alternatively, the $V(z)$ function can expressed using a real argument $x = \sqrt{2{\imath}}\, z$ 
\begin{equation}
\label{eq:V_z_ABCD}
 V(z) = {E(x) \sin \delta(x) \over 1 + \chi} ={1 \over 1 + \chi } {B(x)C(x) - A(x)D(x) \over C^2(x) + D^2(x)} .
\end{equation}
The expressions for functions $A$, $B$, $C$ and $D$ are derived in \citet{Vokrouhlicky1998} and are also provided in Appendix \ref{app:ABCD}.

Apart from the eclipse-induced YS effect, the radiation from the primary to the secondary would cause a so-called ``planetary'' Yarkovsky effect \citep{rub2006,Vokrouhlicky2007}, which replaces the solar radiation with the thermal radiation of the primary in the Yarkovsky effect. Its resulting semimajor axis drift rate can be expressed as 
\begin{equation}
\label{eq:a_pY1}
    \dot a_{\rm pY}  = {2 f_{\rm pY} \mathcal{F_{\rm pY}} \over n }, 
\end{equation}
where
\begin{align}
    f_{\rm pY} & \simeq - { f_{\rm YS} \over c_1 } ,\\
    \mathcal{F_{\rm pY}} & \simeq \mathcal{F} \left({ r_{\rm p} \over 2 a} \right)^2 . \label{eq:F_pY}
\end{align}
Equation~(\ref{eq:F_pY}) results from the fact that the radiation flux from the primary is smaller than the solar radiation flux by a factor of $(r_\p/2a)^2$. It is important that the planetary Yarkovsky effect does not require the eclipse condition and therefore works for high-inclination cases. Noticing that $c_1 \simeq r_\p / \pi a$, we have 
\begin{equation}
\label{eq:a_pY2}
    \dot a_{\rm pY} = -a_\YS {\pi r_\p \over 4 a},
\end{equation}
showing that the planetary Yarkovsky effect is an opposite effect to the YS effect. Considering $r_\p /a < 1$, the YS coefficient not only dominates over the planetary Yarkovsky effect, but also becomes progressively stronger relative to it as the binary semimajor increases.

Combining the eclipse-induced YS effect (Eq.~\ref{eq:a_YS}) and the planetary Yarkovsky effect (Eq.~\ref{eq:a_pY2}), we obtain the total Yarkovsky effect on the binary asteroid
\begin{equation}
    \dot a_\Y = \dot a_\YS + \dot a_{\rm pY} = \dot a_\YS \left(1 - {\pi r_\p \over 4 a}\right).
\end{equation}
For the sake of simplicity, we introduce a combined Yarkovsky coefficient
\begin{equation}
\label{eq:f_Yark}
    f_\Y = f_\YS \left(1 - {\pi r_\p \over 4 a}\right)
\end{equation}
such that the total Yarkovsky effect has a generic form
\begin{equation}
\label{eq:a_Yark}
    \dot a_\Y = {2 f_\Y \mathcal{F} \over n}.
\end{equation}

\subsection{Discussion}
\label{sec2.2}
The sign of the Yarkovsky-induced drift rate $\dot a_\Y$ is the same as the Yarkovsky coefficient $f_\Y$, while the latter is a complicated function that depends on the properties of the binary system (see Eq.~\ref{eq:f_YS_d} and Eq.~\ref{eq:f_YS_s}). Roughly said, for prograde secondaries ($\varepsilon < 90^\circ$ ), the Yarkovsky effect tends to drive the secondary towards the synchronous orbit $a_\syn$ determined by $n = \omega$, while for retrograde secondaries ($\varepsilon > 90^\circ$), the Yarkovsky effect always drives the secondary outward until it leaves the system. The Yarkovsky coefficient $f_\Y$ could have a simpler form in the fast spin regime ($\omega \gg n$) or in the slow spin regime ($\omega \ll n$), as discussed in Appendix~\ref{app:special_cases}.

\begin{figure}
    \centering
    \includegraphics[width = 0.5\textwidth]{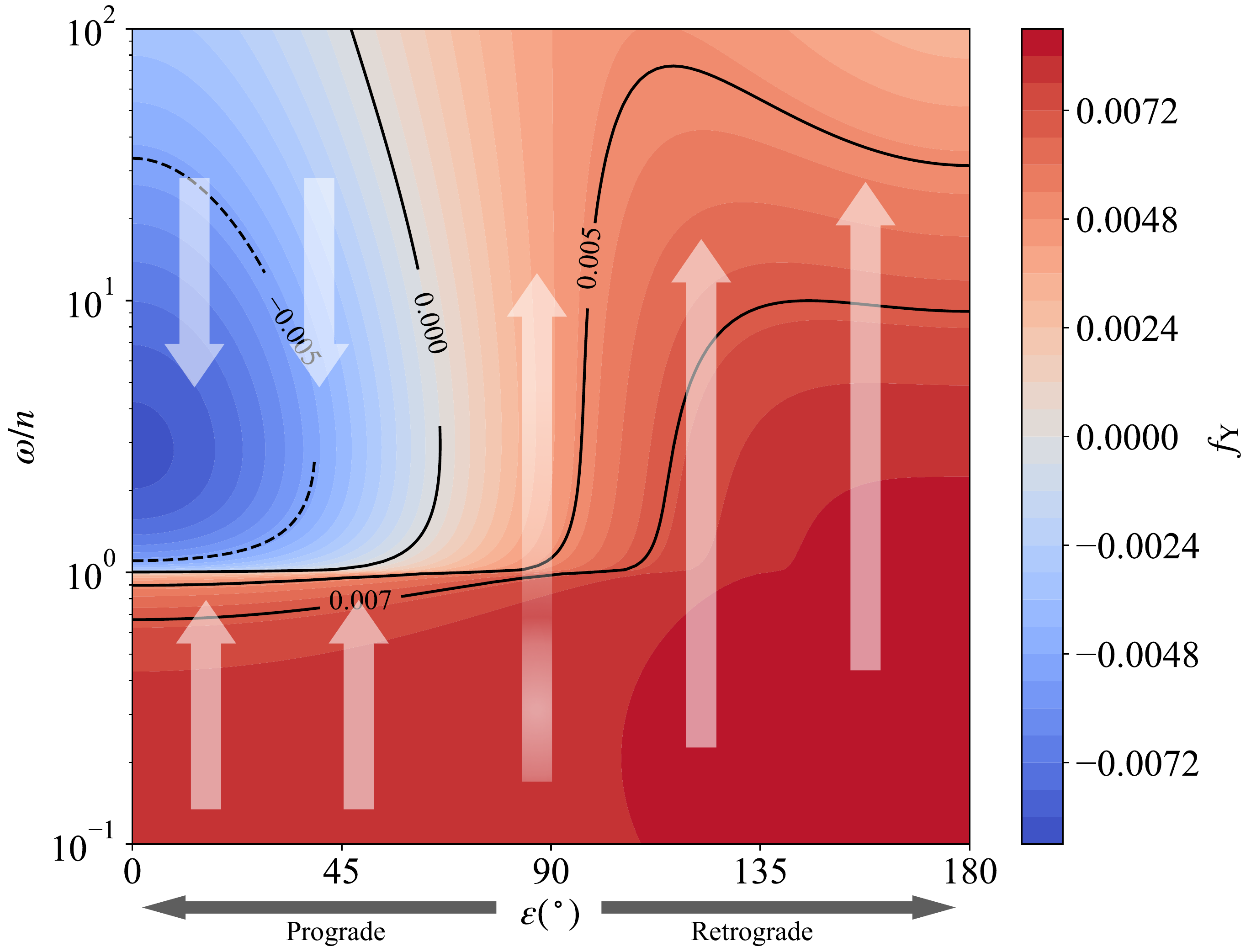}
    \caption{The Yarkovsky coefficient, $f_\Y$, as a function of $m$ and $\varepsilon$ where $m = \omega/n$ is the ratio of the spin frequency to the orbital frequency and $\varepsilon$ is the angle between the spin vector and the orbital vector. The sign of the $f_\Y$ is the same as the sign of $\dot a_\Y$ and $f_\Y = 0.005$ corresponds to $\dot a = 1.4$~cm~yr$^{-1}$ for this system. The direction of the white arrows denotes the evolution direction under the Yarkovsky effect. For small values of $\varepsilon$, the Yarkovsky effect drives $m$ towards 1, otherwise the satellite is driven away from the primary.}
    \label{fig:m_epsilon}
\end{figure}

For the purpose of an illustration, we consider a binary system on a heliocentric circular orbit and $a_{\rm h}=1$~au. We set $r_\p = 1000\m$, $r_\rms = 200 \m$, and the physical parameters $K_\p = K_\rms = 0.1$~W~m$^{-1}$~K$^{-1}$,  $C_\p = C_\rms = 550$~J~K$^{-1}$~kg$^{-1}$ and $\rho_\p = \rho_\rms = 2000$~kg~m$^{-3}$. 
The semimajor axis $a$ of the binary components is fixed at $2650\m$, which corresponds to an orbital period of $10$ hours. As a result, different values of $m$ are obtained solely by changing the spin frequency $\omega$. We also set the obliquity $\varepsilon$ as a free parameter. Figure~\ref{fig:m_epsilon} shows the Yarkovsky coefficient $f_\Y$ as a function of $m$ and $\varepsilon$. Clearly, the Yarkovsky effect drives the secondary orbit to evolve towards the synchronous state ($m=1$) for prograde rotators, but always pushes the retrograde rotators outward. In the blue zone, the Yarkovsky effect maximizes at a spin period of ${\sim}3$ hours, corresponding to the thermal parameter value $\Theta_\omega{\sim}\sqrt{2}$; here $\Theta_\omega = \Gamma \sqrt{\omega}/(\epsilon\sigma T_{\rm sub}^3)$ with the surface thermal inertia $\Gamma=\sqrt{K_{\rm s} C_{\rm s} \rho_{\rm s}}$.

In the most common case seen for the observed binary systems, namely $\varepsilon {\sim}0^\circ$, the Yarkovsky coefficient simplifies (with only the diurnal component contributing)
\begin{equation}
    f_\YS = {4 c_1 \over 9} V(z_{m-1}).
\end{equation}
We introduce the relative angular frequency $\Delta = \omega - n$ , such that $z_{m-1} = \sqrt{-{\imath}} r_\rms / l_\Delta$ where $l_\Delta$ is defined as
\begin{equation}
    l_\Delta = \sqrt{K_\rms \over \rho_\rms C_\rms |\Delta|} =2\, {\rm cm} \left({ |\Delta| \over 2 \times 10^{-4} ~{\rm rad~s}^{-1}} \right)^{-1/2}
\end{equation}
(we use the above-given physical parameters of the satellite surface).
In the case of large bodies $r_s \gg l_\Delta$, readily fulfilled in cases of interest, we can further apply the approximate expression for $V(z_{m-1})$ function
\begin{equation}
 V(z_{m-1}) = -{\Theta_{\Delta} \over 2 + 2\Theta_{\Delta} + \Theta_{\Delta}^2}\, {\rm Sign}(\omega-n) \label{eq:planar}
\end{equation}
with the thermal parameter $\Theta_{\Delta}$ defined as
\begin{equation}
\label{eq:Theta_Delta}
    \Theta_\Delta = {\Gamma \sqrt{|\Delta|} \over \varepsilon \sigma T_{\rm sub}^3}. 
\end{equation}
Note that $f_\Y$ does not depend on the size of the secondary when $r_\rms \gg l_\Delta$. In the regime of $r_{\rms} < l_\Delta$, $f_\Y$ gets smaller when $r_\rms$ decreases. The Yarkovsky coefficient $f_\Y$ depends on the semimajor axis following $f_\Y \propto {r_\p} (1 - {\pi r_\p / 4a})/a$, considering a constant $\Delta$.
As the secondary asteroid is always outside the Roche limit ($a \gtrsim 1.5r_\p$), $f_\Y$ decreases with $a$. 

\begin{figure}
    \centering
    \includegraphics[width = 0.5\textwidth]{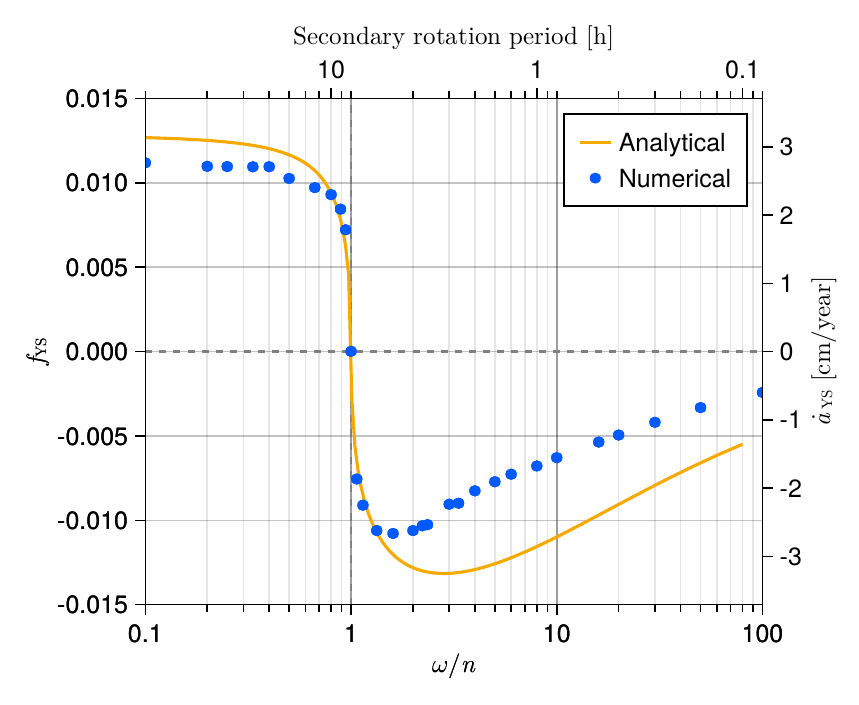}
    \caption{The YS coefficient $f_\YS$ as a function of the ratio of the spin velocity to the mean motion, $\omega / n$. The blue dots show the numerical results of the zero-obliquity case, compared with the analytical solution (orange). The secondary x-axis and y-axis indicate the corresponding rotation period and semimajor axis drift for the given binary system.}
    \label{fig: YS_numerical}
\end{figure}

\subsection{Comparison with numerical simulation}\label{nums}

To validate our analytical solution, we compare our results with numerical solutions. We performed thermophysical simulations using the \texttt{AsteroidThermoPhysicalModels.jl} library, one of the functionalities of the asteroid dynamical simulator \texttt{Astroshaper} (\url{https://github.com/Astroshaper}). This package was originally developed to predict the YORP effect on asteroid 162173 Ryugu, a target asteroid of Japan's Hayabusa2 spacecraft \citep{Kanamaru2021}. The thermophysical model, originally formulated for a single asteroid, has been generalized to include all relevant thermal effects in a binary system. Most importantly, we account for the mutual shadowing between the binary components.

For comparison with the analytical solution, we numerically evaluated the YS-coefficients $f_\YS$ for different orbit and spin periods of the secondary. We consider a binary asteroid with the same parameters as Sec.~\ref{sec2.2}. The binary at $1$ au heliocentric distance has been given zero eccentricity (both the heliocentric and mutual orbits). The spherical binary components were approximated using a triangulated model with $2562$ vertices and $5120$ facets for both the primary and the secondary. The obliquity of the secondary's spin pole is set to be $\varepsilon=0^\circ$.

We performed thermophysical simulations for $100$ thermal cycles to reach converged values of $f_\YS$, with the least common multiple of the secondary's orbit period and the spin period as one cycle. The orbital period of the satellite was fixed at 8 hours, and the rotation period was varied to simulate cases of different $\omega / n$. The radiation flux between the primary and the secondary was hereby ignored to save computational time. At each time step, we calculated the temperature distribution of the asteroids and the thermally induced force on each surface facet, as described in \cite{Rozitis2012}. The thermal force in an asteroid-fixed frame was then transformed into an inertial frame to calculate the acceleration on the secondary. The effective YS coefficients $f_\YS$ are plotted in Fig.~\ref{fig: YS_numerical} as a function of the ratio of the spin velocity to the mean motion, $\omega / n$. The numerical results are in reasonable agreement with the analytical solution (Eq.~\ref{eq:planar}), given its simplicity, providing its justification. A complete parameter survey by the numerical thermophysical model will be presented in future work.

\section{Implications}
\label{sec3}
\subsection{Synchronization of the secondary component}
The majority of the binary asteroid systems are observed to have a synchronized secondary. Simulation of rotational disruption of asteroids shows that the secondary could be born either asynchronous or synchronous due to the frequent reshaping near the Roche limit \citep{Agrusa2024}. Currently, there are two known mechanisms for the synchronization of the secondary asteroid: the tidal effect and the YORP effect. Both of these two effects synchronize the secondary by changing the rotation of the secondary until it gets tidally locked. The tidal bulge raised on the secondary by the primary causes a torque that tends to remove the difference between the spin frequency and orbital frequency. The estimate of the characteristic timescale related to the tidal torque makes use of \citep{Murray1999} 
\begin{equation}
    \dot \omega = { 5 \pi G \rho_\p^2 r_\p^6 \over r_\rms^5 a \rho_\rms} \left( {r_\rms \over a} \right)^5 {k_\rms \over Q}
\end{equation}
with $k_\rms = 0.038\, G \rho_\rms^2 r_\rms^2/\mu$ \citep{Burns1973,Murray1999,Quillen2022} assuming a monolithic structure. While we note that $k_\rms/Q$ is a more fundamental parameter for the tidal effect, we use the parameter $\mu Q$ for ease of comparison with previous work. However, it is important to note that the expression of the tidal love number $k_2$ for rubble piles is still under debate and poorly constrained \citep{Burns1973, Yoder1982, Taylor2011, Jacobson2011b, Dellagiustina2024}. The timescale
$\tau_{\rm t, spin}\simeq \omega/{\dot \omega}$ reads
\begin{equation}
\begin{aligned}
\label{eq:tau_tide}
    \tau_{\rm t, spin} & = { \omega Q \rho_\rms \over 5\pi k_2 G \rho_\p^2  } \left( {a \over r_\p} \right)^6  \\
    & \simeq 10~ {\rm Myr} \left( {8.7 ~{\rm h} \over P_\rms}\right) \left( {a/r_\p \over 2.5 }\right)^6 \left( {r_\rms \over 0.2 \km}\right)^{-2} \left( {\mu Q \over 10^{11} {\rm ~Pa}}\right),
\end{aligned}
\end{equation}
where $P_\rms=2\pi/\omega$ is the satellite rotation period.
The value of $\mu Q$, which varies by a few orders of magnitude in the literature \citep{Burns1973, Goldreich2009, Efroimsky2015, Caudal2023, Pou2024}, is still uncertain for rubble piles and its dependence on the size is also poorly known. 

The radiative torque due to the irregular shape, namely the YORP torque, can spin up or down the secondary. The direction of the YORP torque, which depends on the shape and rotation state, is still poorly understood. The timescale of the YORP effect is \citep{Rubincam00, Bottke06, Marzari2020}
\begin{equation}
\label{eq:tau_YORP}
    \tau_\YORP \simeq 42 {~\rm kyr} \left( {r_\rms \over 0.2 \km } \right)^2 \left( {8 ~{\rm h} \over P } \right) \left( {a_{\rm h} \over 1~{\rm au} } \right)^2.
\end{equation}
It is obvious that the YORP timescale is much shorter than the tidal timescale, implying that the YORP effect could be the major mechanism for the synchronization of the secondary asteroid. However, there are two issues. First, the YORP effect may be highly sensitive to the fine-scale surface irregularities  \citep{Statler2009, Breiter2009, Cotto2015}. In most cases, this information is beyond the resolution of the available observations. As a result, predicting even the instantaneous YORP value may be very difficult \citep[this is the reason why the presently achieved YORP detections are often smaller than theoretically expected; see the latest compilation in the discussion section of][]{Durech2024}. From a long-term perspective, the movement of boulders \citep{Golubov12} and formation of impact craters \citep{Zhou2022, Zhou2024} could modify or even reverse the direction of the YORP torque. Building on the YORP's shape sensitivity, \cite{Bottke2015} introduced the ``stochastic YORP'' concept and showed it overall weakens the long-term effects of nominal YORP. Second, theoretically, the YORP torque has an equal probability of taking a positive or negative sign. Therefore, we would expect that half of the secondaries are asynchronous due to the wrong direction of the YORP torque (i.e. opposite to $n - \omega$), which is inconsistent with the observed dominating synchronous population {with tight orbits \citep[e.g. $a < 2.2 r_\p$ or $P_{\rm orb} < 20~$h, see][]{Pravec2016}. Asynchronous secondaries occur more frequently in wide orbits compared to tighter ones \citep{Pravec2016}, implying a correlation to some mechanism that influences orbital configurations.} 

While there is no clear answer for the dominating mechanism of synchronization of the secondary, we find that the Yarkovsky effect also drives the prograde rotators towards the synchronous orbit (Sec.~\ref{sec2}). Here, we {estimate} the timescale for the Yarkovsky effect to synchronize the orbit:
\begin{equation}
\begin{aligned}
\label{eq:tau_Y}
    \tau_{\Y} &= {a \over \dot a_\Y} = {a n \over 2 f_\Y \mathcal{F}} \\
    & \simeq 160 {\rm kyr} \left( r_\p \over 1\km \right) \left( r_\rms \over 0.2\km \right) \left( a/r_\p \over 2.5  \right)^{1/2} \left( 0.005 \over f_{\rm Y,0} \right) \left( a_{\rm h} \over 1{\rm ~au} \right)^2,
\end{aligned}
\end{equation}
{where $f_{\rm Y,0}$ is the Yarkovsky coefficient at $a = 2.5 r_\p$}. {The Yarkovsky effect synchronizes the secondary by principally modifying its orbit, while the tidal and the YORP effects change its spin rate.} {Compared to the YORP timescale (Eq.~\ref{eq:tau_YORP}),} Eq.~(\ref{eq:tau_Y}) suggests that the Yarkovsky effect could operate more efficiently than the stochastic YORP {for relatively large objects. The Yarkovsky could also dominate over the tidal effect unless the satellite is close-in or large, or the value $\mu Q$ is smaller than assumed.} Therefore, we propose that the Yarkovsky effect could be---at least in some small binaries---the major mechanism to synchronize the secondary.
Let us give two examples of interest, namely Didymos-Dimorphos and Dinkinesh-Selam systems.

By abruptly reducing the binary orbit period, the NASA DART impact may have broken Dimorphos from synchronous rotation. Due to the oblate shape of Dimorphos \citep{Daly2023}, the spin-orbit coupling could be very weak and instead of Dimorphos' long axis librating about the post-impact synchronous state, it could be circulating \citep{Richardson2023}. If this is the case, we can estimate the present Yarkovsky drift rate. By setting the orbital period to be $11.37~$hours and the spin period $12~$hours, we obtained an estimate of $\dot a {\sim}7.6~$cm~yr$^{-1}$, which could be examined by the subsequent space mission ESA Hera that will launch in October 2024 to visit Didymos in fall 2026 \citep{Michel2022}. However, we note that this possibility is only one of many possible Dimorphos's post-impact spin states {including tumbling} \citep{Agrusa2021, Richardson2023}. {The Yarkovsky effect does not vanish for tumbling objects (not even in the strong tumbling regime such as the long axis mode). For a weak tumbling regime (short axis mode), the Yarkovsky effect could be acceptably well represented using a traditional formulation (with rotation about the principal axis of the inertia tensor) and assuming: (i) spin axis oriented along the rotational angular momentum, (ii) rotation period close to the precession period, and (iii) the shape given by convex hull swept during the tumbling cycle \citep{Vokrouhlicky2015b}. There are many examples of tumbling NEAs, such as 99942 Apophis \citep{Pravec2005, Pravec2014, Vokrouhlicky2015b, Del2018, Perez2022} and 4179 Toutatis \citep{Vokrouhlicky2005b, Chesley2015, Del2018}, for which the Yarkovsky signal was firmly detected, pretty much as expected and within the expected range of non-tumbling state. Thus, we suspect that the rule of the Yarkovsky effect would still be valid for tumbling components in a binary system, while a more thorough investigation is required for confirmation in the future. There is some observational evidence indicating Dimorphos may be in some excited tumbling states \citep{Pravec2024}, where the satellite's  longest axis is approximately tidally locked to the direction towards the primary. Given its on-average synchronous rotation, the Yarkovsky effect could be weak or even shut off in this case, but more detailed analysis of this interesting system is needed.}

In the latter case, the Dinkinesh-Selam binary recently discovered by the Lucy mission \citep{Levison2024}, the secondary asteroid Selam appears to be synchronous with a wide orbit at $a/r_\p \simeq 9$. Selam is likely a contact binary, possibly formed by the merger of two satellites. If this is the case, Selam was unlikely to be in synchronous rotation following a merger, requiring some synchronization mechanism to explain its present spin state. The timescale for the tidal despinning could be as long as $\sim 3~$Gyr according to Eq.~(\ref{eq:tau_tide}), due to the wide orbit. However, the typical collisional lifetime of asteroids with the size of Dinkinesh \citep[$0.8$~km in diameter][]{Levison2024} is about $0.3$~Gyr \citep{Bottke2005}. Since this lifetime is much smaller than the tidal despinning timescale, it is therefore unlikely that Selam was synchronized by the tidal effect, while the Yarkovsky effect can synchronize the orbit quickly. We found the Yarkovsky timescale could be $\sim 1~$Myr, by setting $r_\p = 720$~m, $r_\rms = 277$~m, $a = 3.1~$km, $\rho_\p = \rho_\rms = 2.4~$g~cm$^{-3}$, $a_{\rm h} = 2.19~$au and $e_{\rm h} = 0.11$. Therefore, we propose that the Yarkovsky effect could be the main reason for its current synchronous state.


\subsection{Long-term evolution of binary asteroids}

Let us now consider possible evolutionary pathways of small binary systems in general terms, extending the canonical view (with tides and BYORP operating) by the Yarkovsky effect. Assume the parent body of the binary is disrupted either by rotational fission or a catastrophic collision. The resulting fragments that are bound to the parent body accumulate to form a satellite, which can be either in synchronous or asynchronous rotation \citep{Agrusa2024}. The synchronous secondary evolves under the tidal and BYORP effects, resulting in a final state at the tide-BYORP equilibrium location (given that the BYORP torque is negative) or in a migration outward until it leaves the system or becomes chaotic (if the BYORP torque is positive) \citep{Cuk2007, Cuk2010, Jacobson2011b, Jacobson2014}. Assuming $a_{\rm B-t}$ is the location of the BYORP-tide equilibrium where $\dot a_\BYORP = -\dot a_\tide$, secondaries with a large BYORP coefficient $f_\BYORP$ such that $a_{\rm B-t} < a_\Roche$ will cross the Roche limit and get tidally disrupted \citep{Cuk2010}.

If the secondary is born asynchronous or perturbed into such a rotation state, it will evolve under the joint effect of tides and the Yarkovsky effect. If it happens to be in a spin-orbit resonance, BYORP is also active \citep{Jacobson2014}. For retrograde rotators, both tides and the Yarkovsky effect expand the secondary's semimajor axis until it is lost, forming an asteroid pair. The timescale for this process is $\min(\tau_\tide, \tau_\Y) \lesssim \tau_\Y$, considering the strength of tides declines rapidly with the semimajor axis. In this case, the two components of the asteroid pair have opposite spin poles, different from asteroid pairs produced by rotational fission \citep{Pravec2010,Pravec2019} or BYORP \citep{Cuk2007}. For prograde rotators, the Yarkovsky effect will shrink the orbit if the secondary spins faster than the mean motion. If the secondary has a spin period $P$ shorter than the orbital period at the Roche limit $P_\Roche$, which is given by
\begin{equation}
    P_\Roche = 2\pi \sqrt{3 \over  4\pi G \rho_\p} \left( {a_\Roche \over r_\p} \right)^{3/2} \simeq 4.3 ~{\rm hr} \left( {a_\Roche \over 1.5r_\p} \right)^{3/2},
\end{equation}
then the secondary will migrate to this point and undergo significant reshaping or even tidal disruption. The typical Roche radius is given $\sim 1.5~r_\p$ \citep{Holsapple06}, assuming the binary components have equal density. The traditional hydrostatic Roche radius is roughly 2.46~$r_\p$, 
corresponding to an orbital period of $\sim 9~$h. 
The distribution of orbital periods of known binary systems shows a cut-off at 11 hours \citep{Pravec2006}. 
More recently confirmed binaries have confined the cut-off orbital period to 10.5 hours, suggesting the Roche radius could be 2.7~$r_\p$ for rubble piles with a weak structure or low density, if the cut-off in the orbital period distribution is caused by tidal disruption. If the secondary has sufficient material strength (e.g. small monoliths), it is also possible that it would continue migrating inwards and form a contact binary. Note that this process requires the Yarkovsky effect to overcome the outward torque due to tides. The semimajor axis $a_{\rm Y-t}$ where tides balance the Yarkovsky effect can be obtained by equating Eq.~(\ref{eq:a_tide}) to 
Eq.~(\ref{eq:a_Yark})
\begin{equation}
    a_{\rm Y-t} = 0.84\, r_\p \left( { r_\rms \over 0.2\km} \right)^{4/7} \left( { 0.005 \over f_\Y} \right)^{1/7} \left( { 10^{11} {\rm Pa} \over \mu Q} \right)^{1/7}.
\end{equation}
This implies that the Yarkovsky effect easily overcomes tides for small binaries unless the tidal effect is much stronger than assumed. Changing $f_\Y$ to $f_\BYORP$ makes the above equation for $a_{\rm B-t}$. Considering the uncertainty of the tidal effect, we assume that $a_{\rm B-t}$ and $a_{\rm Y-t}$ could be located either inside or outside the $a_\Roche$. If $a_{\rm Y-t} > a_\Roche$, the secondary will be stopped outside the Roche limit and be in a Yarkovsky-tide equilibrium state. However, this state can only last for a YORP timescale or a tidal despinning timescale, as the YORP torque or tides change the spin to shut off the Yarkovsky effect. For rotators with $P > P_\Roche$, the Yarkovsky effect will move the secondary towards the synchronous orbit $a_\syn$, except in the special case where the Yarkovsky effect is negative and balanced by tides, leading to a temporary Yarkovsky-tide equilibrium state at $a_{\rm Y-t}$.

To summarize, we provide some predictions on the secondary dynamical state based on the Yarkovsky effect: (1) retrograde secondaries should be relatively far from the primaries; (2) some asteroid pairs could have opposite spin directions if they are formed by the Yarkovsky effect; (3) secondaries with a period shorter than {the orbital period at Roche radius, ranging from $\sim 4$ to $\sim 10$ hours}, should be destroyed in a Yarkovsky timescale ($\sim 0.1~$Myr), otherwise they should be in a Yarkovsky-tide equilibrium (asynchronous) that lasts for a YORP timescale; (4) secondaries with a period longer than {Roche orbital period} should become synchronous after a Yarkovsky timescale.

\begin{deluxetable*}{rlclllllllcllcl}[t] 
 \tablecaption{\label{tab:real_asteriod}
  The Yarkovsky effect predicted for known asynchronous binaries}
 \tablehead{
   \colhead{Name} & \colhead{$a_{\rm h}$ [au]} & \colhead{$e_{\rm h}$} &
   \colhead{$r_\p$ [km]} & \colhead{$r_\rms/r_\p$} & \colhead{$a/r_\p$} & \colhead{$P_{\rm orb}$ [h]} & \colhead{$P_\rms$ [h]} & \colhead{$f_\Y\; (\times 10^{-3})$} &
   \colhead{$\dot a_\Y$ [cm~yr$^{-1}$]}
 }
\startdata
 (1509) Esclangona & 1.866 & 0.032 & 4.25 & 0.33 & 49.2 & 768 & 6.6422 & -0.54 / +0.41 & -0.52 / +0.39 \\ 
 (2486) Metsahovi & 2.269 & 0.08 & 4 & 0.30 & 18.3 & 172.6 & 2.64 & -0.97 / +1.0 & -0.17 / +0.17\\ 
 (2623) Zech & 2.255 & 0.234 & 3.4 & 0.29 & 14.1 & 117.2 & 18.718 & -2.0 / +1.7 & -0.29 / +0.25\\
 (32039) 2000 JO23& 2.223 & 0.283 & 1.3 & 0.65 & 33.1 & 360 & 11.09 & -0.81 / +0.67 & -0.45 / +0.37\\ 
 (311066) 2004 DC & 1.634 & 0.400 & 0.15 & 0.20 & 4.6 & 23 & 7 & -5.3 / +4.6 & -10.2 / +9.0\\ 
\enddata
\tablecomments{The orbital period of the binary components relative to each other is $P_{\rm orb}=2\pi/n$, while the rotation period of the secondary is $P_\rms = 2\pi/\omega$. The Yarkovsky coefficient $f_\Y$ and the drift rate $a_\Y$ are calculated for an obliquity equal to $0^\circ/180^\circ$. The thermal parameters are assumed to be the same as those used in Sec.~\ref{sec2.2}. {It is important to note that for the distant satellites, the Yarkovsky effect could be diminished (see Sec.~\ref{sec3.3}).}}
\end{deluxetable*}

\subsection{Predicted orbital drift rates of real binary asteroids}
\label{sec3.3}
As of now, there are 66 binary asteroid systems with documented secondary spin periods \citep{Pravec2007, Warner2009, Pravec2012, Pravec2016, Monteiro2023}. Among these, {five} are known to exhibit spin periods that are different from their orbital periods up to date. {While there are other potentially asynchronous binary systems, their information is either undetermined or incomplete \citep{Pravec2016}.} Consequently, we computed the theoretical Yarkovsky drift rates for these {five} asynchronous binary asteroids for future tests.

It is important to note that the Yarkovsky effect remains applicable to those binary systems with $n = \omega$, provided their obliquity is non-zero. However, due to insufficient data on the spin vectors, The information regarding the axial tilt of real asteroids remains ambiguous. Here our focus is limited to asteroids with $n \neq \omega$, for which BYORP does not work. We estimated the Yarkovsky effect for these bodies in the limiting cases $\varepsilon = 0^\circ$ and $180^\circ$. The result is shown in Table~\ref{tab:real_asteriod}. We notice that there are a few asteroid binaries (e.g. Esclangona, Arlon) with a large separation, although their spin periods are much smaller than the orbital periods ($m \ll 1$). This large separation was explained as a result of the BYORP-induced expansion and subsequent desynchronization \citep{Jacobson2014}. We found that the Yarkovsky effect could also expand the mutual orbit, if the secondaries have obliquities around or larger than $90^\circ$, which is left for future observational tests. {The Yarkovsky effect decreases with increasing separation, not only because of the decreasing time fraction in the shadow over a mutual orbit (i.e. $c_1$ in Eq.~\ref{eq:f_YS_d} and \ref{eq:f_YS_s}), but also due to the challenges in maintaining the shadow condition over the heliocentric orbit. For a distant satellite, the inclination of its orbit about the primary should be confined in a narrower interval of values than for close satellites. Accounting for the potential non-existence of the shadow in a relatively high-inclination heliocentric orbit, the $c_1$ should be revised as the time fraction that the secondary spends in the shadow over a heliocentric orbit \footnote{This reduces to $c_1 \simeq r_\p / \pi a$ when $i = 0$.}. Therefore, thorough orbital modeling is crucial for assessing the Yarkovsky drift rates for distant satellites. The predictions provided by Table ~\ref{tab:real_asteriod} are based on the simplest situations and therefore only give the upper limits of Yarkovsky drift rates.}

There are also some binaries (e.g. 1994 CC, 2001 SN263, 2004 DC etc.) with large heliocentric eccentricities, therefore they may also suffer from strong planetary perturbations that modify the mutual orbit and the spin state. Since the Yarkovsky effect is much more deterministic than the BYORP effect once the rotational state is known, our hypotheses can be easily examined by future observations.

\section{Conclusions}
\label{sec4}
The Yarkovsky effect is the radiative force acting on a rotating object with non-zero thermal inertia, gradually altering its orbit over the long term. In this work, we investigate the Yarkovsky effect on a binary asteroid system. The binary Yarkovsky effect, manifesting primarily on the secondary asteroid, comprises two main components: the Yarkovsky-Schach effect and the planetary Yarkovsky effect. The former is the net Yarkovsky force averaged over the mutual orbital period, due to the eclipse caused by the primary on the secondary. As a result of the eclipse condition, the Yarkovsky-Schach effect is only significant for low-inclination binary asteroids (e.g. $i < r_\p/a$), which are common in both the NEA and MBA populations. The planetary Yarkovsky effect is simply produced by the radiation from the primary asteroid, instead of the Sun. For low-inclination asteroids, the Yarkovsky-Schach effect dominates over the planetary Yarkovsky effect. The direction of the binary Yarkovsky effect depends on the obliquity and the difference between the spin rate and the mean motion of the secondary, while the magnitude depends on the thermal and orbital properties of the binary system. In general, the binary Yarkovsky effect moves the secondary to make the mean motion match the spin rate on a timescale of $\sim 0.1~$Myr.

We found that for prograde-rotating secondaries, the Yarkovsky effect can synchronize the secondary (i.e. $\omega = n$) by orbit modification, on a timescale much shorter than tidal despinning, {except for large or close-in secondaries}. {On the other hand, the YORP effect could 
be more efficient for synchronization of small secondaries. This is because of its timescale depending on $\sim r_{\rm s}^2$ rather than $\sim r_{\rm s}$ for the Yarkovsky effect.} This brings us new insights about the mechanism of the synchronization of binary asteroids and the underlying reason why the majority of binary asteroids are found to be in synchronous states. Our calculations also predict that the secondary asteroids with spin periods shorter than the orbital period around the Roche limit {(e.g. from $\sim 4$ to $\sim$10 hours)} will fall into the Roche limit quickly driven by the Yarkovsky effect and then get tidally disrupted, reshaped or accreted on the primary. In addition, some asynchronous binaries might be in the Yarkovsky-tide equilibrium state where the orbit does not drift, but such a state may be quickly broken by the YORP effect or tides. For retrograde secondaries, the Yarkovsky effect would drive them outward until they leave the binary system due to planetary perturbations or collisions, producing asteroid pairs. In this scenario, the two components of the asteroid pair would exhibit opposite spin directions.  

We also calculated the Yarkovsky-caused drift rate for known asynchronous binaries, listed in Table~\ref{tab:real_asteriod} for future observational tests. Some of the asynchronous binary asteroids have wide mutual orbits, which could be the result of the Yarkovsky effect on retrograde secondaries. We found that the synchronization of the Dinkinesh-Selam system discovered by the Lucy spacecraft could be due to the Yarkovsky effect, considering that tides are weak for such a distant secondary. In addition, we calculated the possible Yarkovsky effect on Didymos-Dimorphos system in its state following the impact of the NASA DART mission, which might have perturbed it into an asynchronous state. The Yarkovsky coefficient $f_\Y$ is around 0.0067 and the resulting semimajor axis drift rate is $\dot a \simeq 7.6~$cm~yr$^{-1}$. This could be examined by in-situ observation conducted by the space mission ESA Hera during its rendezvous with Didymos in late 2026.

\vfill\null
\newpage

\section{Acknowledgments}
We thank Miroslav Bro\v{z} and Jay McMahon for useful discussions that helped improve the submitted manuscript. We thank the anonymous reviewer for a referee report that significantly improved the paper. We acknowledge support from the Universit\'e C\^ote d'Azur. Wen-Han Zhou would like to acknowledge the funding support from the Chinese Scholarship Council (No.\ 202110320014). 
The work of David Vokrouhlick\'{y} and Petr Pravec was partially supported by the Czech Science Foundation (grant~23-04946S).
Masanori Kanamaru acknowledges the funding support by the JSPS KAKENHI No. JP22J00435/JP22KJ0728 and MEXT Promotion of Distinctive Joint Research Center Program Grant Number JPMXP0622717003/JPMXP0723830458.
Harrison Agrusa was supported by the French government, through the UCA J.E.D.I. Investments in the Future project managed by the National Research Agency (ANR) with the reference number ANR-15-IDEX-01.
Patrick Michel acknowledges funding support from the French space agency (CNES) and the European Space Agency (ESA).

\appendix

\section{Nomenclature}
The symbols used in this paper are listed in the Table.~\ref{tab:Nomenclature}.

\begin{deluxetable*}{cl|cl}[t] 
 \tablecaption{\label{tab:Nomenclature}
  Notation used throughout this paper}
 \tablehead{
  \colhead{Symbols} & & \colhead{Subscripts} & 
 }
\startdata
  $a$     &  semimajor axis of a mutual orbit & $\p$ & Primary component of a binary asteroid \\
  $k$     &  Love number    & $\rms$ & Secondary component of a binary asteroid \\
  $n$     &  Mean motion    & h & Heliocentric parameter \\
  $Q$     &  Quality factor & B & BYORP \\
  $r$     &  Radius & Y & Yarkovsky \\
  $f$     &  Dimensionless coefficient & YS & Yarkovsky-Schach \\
  $A$     &  Bond albedo & pY & planetary Yarkovsky \\
  $\rho$  &  Density & Y-t & Yarkovsky-tide equilibrium \\
  $\mu$   &  Rigidity & B-t & BYORP-tide equilibrium \\
  $i$ &  Inclination & Roche & Roche limit \\
  $\varepsilon$ & Obliquity & & \\
  $\omega$ &  Spin rate of the secondary & & \\
  $\tau$   &  Timescale & & \\
  $\Theta$ &  Thermal parameter & & \\
  $\Gamma$ &  Thermal inertia & & \\
  $\mathcal{F}$ & Nominal radiative force per unit mass & & \\
\enddata
\end{deluxetable*}

\section{Functions $A(x)$, $B(x)$, $C(x)$ and $D(x)$}
\label{app:ABCD}
The functions $A(x)$, $B(x)$, $C(x)$ and $D(x)$, useful to express the V($z$) in real
notation (Eq.~\ref{eq:V_z_ABCD}), are given by \citep[e.g.,][]{Vokrouhlicky1998,Vokrouhlicky1999}
\begin{align}
  A(x) &=  -\left(x+2\right) - e^x \left[\left(x-2\right) \cos x - x \sin x\right],\\
  B(x) & = -x - e^x \left[x \cos x + \left(x-2\right) \sin x\right], \\
  a(x) & = 3\left(x+2\right) + e^x \left[3\left(x-2\right)\cos x + x\left(x-3\right)\sin x\right], \\
  b(x) & = x\left(x+3\right) - e^x \left[x\left(x-3\right) \cos x - 3\left(x-2\right) \sin x\right],
\end{align}
with $C(x)=A(x)+ \chi\,a(x)/(1+\chi)$ and $D(x)=B(x)+\chi\,b(x)/(1+\chi).$

\section{Special cases for $f_\Y$} \label{app:special_cases}

The formula of the Yarkovsky coefficient $f_\Y$ can be simplified in some special cases. In the fast spin regime, where $\omega \gg n$ such that $m\gg 1$, $V(z_{m+1}) \simeq V(z_{m-1}) \simeq V(z_m)$ and $z_m=\sqrt{{-\imath} m}\, r_\rms/l_n$. In this
case, one could replace the penetration depth $l_n$ of the thermal wave at mean motion frequency $n$ with a penetration depth $l_d=l_n/\sqrt{m}$ of the thermal wave at rotation frequency, thence $z_m=\sqrt{{-\imath}}\, r_\rms/l_d$. Since
$\cos^4(\varepsilon/2) - \sin^4(\varepsilon/2) = \cos \varepsilon$, we obtain a simpler form of Eq.~(\ref{eq:f_YS_d}) reading
\begin{equation}
\label{eq:f_YS,d,fast}
    f_{\YS, \rm d}= {4 c_1 \over 9 } V(z_m) \cos \varepsilon .
\end{equation}
This equation resembles the classic Yarkovsky effect for a single asteroid orbiting around the Sun, but is multiplied by $-c_1$. Therefore, the Yarkovsky is maximized when the spin thermal parameter $\Theta_\omega \sim \sqrt{2}$, which is defined as
\begin{equation}
\label{eq:Theta}
    \Theta_\omega = {\Gamma \sqrt{\omega} \over \varepsilon \sigma T_{\rm sub}^3} 
     \sim \sqrt{2}\, \left( {\Gamma \over 200 \J \m^{-2} \K^{-1} \s^{-1/2}} \right) \left( {P_\rms \over 3\h} \right)^{-1/2} \left( {a_\rmh \over 1 \au} \right)^{\frac{3}{2}}.
\end{equation}
It is obvious that when $\varepsilon = 0^\circ$, there is only the diurnal component left, leading to the inward migration of the secondary asteroid and the decrease of $m$ (since $n$ becomes larger).

In the slow spin regime, where $\omega \ll n$ such that $m \sim 0$, we have $V(z_{m+1}) \simeq -V(z_{m-1}) \simeq V(z_1)$ and $z_1=\sqrt{{\rm -\imath}}\, r_\rms/l_n$. Therefore,
\begin{equation}
    f_{\YS, \rm d} = -{4 c_1 \over 9 } V(z)\, \left(1 - {1\over 2}\sin^2 \varepsilon\right) .
\end{equation}
Interestingly, when combined with the seasonal component, we get the total YS coefficient $f_\YS$
\begin{equation}
    f_\YS = -{4 c_1 \over 9 } V(z_1) 
\end{equation}
that is independent of the obliquity $\varepsilon$. Here, $f_\YS > 0$ which leads to an outward migration and increasing $m$ (since $n$ becomes smaller).

In the synchronous regime where $m = 1$, considering $V(z_0) \rightarrow 0 $, we obtain
\begin{equation}
    f_{\YS, \rm d} = {4c_1 \over 9} V(z_{-2})\,\ \sin^4 {\varepsilon \over 2},
\end{equation}
and $f_{\YS, \rm s}$ remains the same. This will result in a zero Yarkovsky effect given that $\varepsilon = 0^\circ$. For the case of $\varepsilon \neq 0^\circ$, the Yarkovsky effect always transfers a positive angular momentum, driving the secondary outward.

\FloatBarrier
\bibliography{references}{}
\bibliographystyle{aasjournal}
\end{CJK*}
\end{document}